# BIOLOGICAL UNIVERSALITY YIELDS NEW KIND OF LAWS

Mark Ya. Azbel',

School of Physics and Astronomy, Tel-Aviv University,

Ramat Aviv, 69978 Tel Aviv, Israel[+]

and

Max-Planck-Institute für Festkorperforschung – CNRS,

F38042 Grenoble Cedex 9, France

Biological approximations, which are universal for diverse species, are well known. With no other experimental data, their invariance to transformations from one species to another yields exact conservation (with respect to biological diversity and evolutionary history) laws, which are inconsistent with known physics and unique for self-organized live systems. The laws predict two and only two universal ways of biological diversity and evolution; their singularities; a new kind of rapid (compared to lifespan) adaptation and reversible mortality, which may be directed. Predictions agree with experimental data, and call for new concepts, insights, and microscopic theory.

PACS numbers: 05.65.+b; 89.75.-k; 87.23.Kg; 87.10.+e

Every new field in science yielded new laws and fundamental constants, e.g.: relativity and the speed of light, quantum mechanics and the Planck constant, statistical mechanics[1] and the Boltzmann constant. Models followed rather than preceded new concepts and laws. This paper proves biological complexity is not an exception: it does not reduce to known physics, and yields new laws and fundamental constants. Naturally, new physics may be unraveled from experiments only. Start with well known empirical relations[2-4] which reduce basal (i.e., resting) oxygen consumption rate and life span for all animals, heartbeat time for animals with heart to animal mass (more details later). Biological data depend on a multitude of unspecified factors in often poorly controllable and reproducible conditions. Yet, the number of basal oxygen molecules, consumed per body atom per maximal life span, is 12 ± 2.5 for all animals[3], whose body mass changes ~ 10 billion times. Within their (relatively high) accuracy, all allometric relations are conserved despite transformations from one species to another, with different biological complexity and evolutionary history. Thus, they present "conservation laws" in biology and evolution. The laws are valid in a living system, which strongly interacts with its non-stationary heterogeneous environment (via metabolism, energy loss, etc). Yet, they do not explicitly depend on time or on any characteristics of the environment. In spite of enormous biological complexity and diversity of animals, the laws reduce to the animal mass only. Such laws are inconsistent with known physics (e.g., energy in physics is conserved in a stationary system only; it includes potential energy, which depends on many variables). They must be related to fine tuned adjustment (of an animal to its biology and environment), which is unique for self-organized live systems[1]. Such adjustment, which provides universal, i.e. biologically non-specific (independent of biology and environment), laws, is a physical and biological challenge. The laws specify perspectives of and impose limitations on biological diversity and its evolutionary changes.

With no other experimental data, the very existence of conservation laws, invariant to a wide class of transformations, is sufficient to predict their exact functional form in a perfect ("canonic") situation. (This is important, since experimental data are limited and relatively inaccurate.) Suppose macroscopic canonic quantities $u$ and $v$ yield an exact universal relation $u = f(v)$. The values of $u$ and $v$ in a population are the averages of their values $u_G$ and $v_G$ in different population groups (e.g., groups in different living conditions). If the distribution function of $v_G$ in $v$ is $c(v,v_G)$, then

$$\int c(v,v_G)dv_G = 1; \quad v = \int v_G c(v,v_G)dv_G \equiv \langle v_G \rangle . \tag{1}$$

So, $u = f(v) = f(\langle v_G \rangle)$. Universality implies that $u_G = f(v_G)$. Suppose $u$ is an additive quantity, i.e. $u = \langle u_G \rangle = \langle f(v_G) \rangle$. Then $\langle f(v_G) \rangle = f(\langle v_G \rangle)$, i.e.

$$\int c(v,v_G) f(v_G) dv_G = f\left[\int c(v,v_G) v_G dv_G\right] \tag{2}$$

Equation (2) is a functional equation which is linear in $f$ and non-linear in $c$. Consider a special case of

$$c = c_1 \delta(v_G - w_1) + c_2 \delta(v_G - w_2), \text{ where } c_1 + c_2 = 1. \tag{3}$$

Then, by Eq. (2), $c_2 f'(w_1) = c_2 f'(w_2)$. Thus, either

$$f(v) = av + b, \tag{4a}$$

where $a$ and $b$ are constants, or $c_2 = 0$, i.e.

$$c(v,v_G) = \delta(v_G - v) \tag{4b}$$

A special case (3) implies that Eqs. (4a, 4b) are necessary for $f(v)$ being a solution to Eq. (2). Since in a general case Eqs. (4a) and (4b) satisfy Eq. (2), they are sufficient for $f(v)$ to be a solution to Eq. (2). Thus, Eqs. (4a, 4b) present its general solution.

Equation (4b) poses an unusual physical and biological challenge. The value of $v_G$ depends on a set $\eta$ of multiple biological factors, thus Eq. (4b) implies that $v_G(\eta) = v$ is constant inside the corresponding multidimensional $\eta$ manifold. An infinitesimally close $v + dv$ corresponds to a different manifold $\eta'$, which is infinitesimally close to $\eta$ along a certain area (e.g., in a toy one dimensional case $v_G(\eta) = v$ when $\eta_1 \leq \eta \leq \eta_2$, and $v_G(\eta) = v + dv$ when $\eta_2 + d\eta_2 \leq \eta \leq \eta_3$). For a continuous $v$ this implies a singularity of $v_G$ at its $\eta$ boundary, i.e. at every value of $v$. Finite number of manifolds (e.g., species) "quantizes" continuous $v$ into (evolutionary metastable) constants. Naturally, singularities at all points and their "quantization" are inconsistent with and a challenge to any theory.

Equations (4a) and (4b) present a general solution with no singularities in $f(v)$. Allow for $f(v)$ singularities at points $v^{(j)}$ ($j = 1, 2, \ldots$), i.e. consider the possibility of transitions between different universal solutions. Validity of the solution (4a) is related to its linearity and to the normalization condition (1). Singular points $v^{(j)}$ determine successive universal intervals $(v^{(j)}, v^{(j+1)})$ where a canonic population is distributed, i.e. where

$$\int_{v^{(j)}}^{v^{(j+1)}} c(v, v_G) dv_G = 1 \tag{5}$$

There, similar to the previous case, a piecewise linear law

$$u = a^{(j)} v + b^{(j)} \text{ if } v^{(j)} \leq v \leq v^{(j+1)}, \tag{6}$$

where $a^{(j)}$ and $b^{(j)}$ are universal constants, is a universal solution to Eq. (2). At any intersection $v^{(j)}$, Eq. (5) implies

$$c(v^{(j)}, v_G) = \delta(v_G - v^{(j)}). \qquad (7)$$

Equations (5) and (7) yield a "heterogeneity exclusion principle" in a canonic population – they exclude segments outside a given interval and any heterogeneity at the segment boundaries (but impose no limitations on heterogeneity inside each of the segments). Since different groups are heterogeneous in their complete sets of factors $\eta$, Eq. (7) implies vanishing susceptibility of $v$ to $\eta$ at $v^{(j)}$.

Equation (6) may be presented in the form:

$$u = \xi^{(j)} u^{(j)} + \xi^{(j+1)} u^{(j+1)}; \; \xi^{(j)} = \frac{v - v^{(j)}}{v^{(j+1)} - v^{(j)}}; \; \xi^{(j+1)} = 1 - \xi^{(j)}, \qquad (8)$$

(where $v^{(j)} \leq v \leq v^{(j+1)}$). Equation (8) maps Eq. (6) onto the coexistence of phases $u^{(j)} = f(v^{(j)})$ and $u^{(j+1)} = f(v^{(j+1)})$ with the "concentrations" $\xi^{(j)}$ and $\xi^{(j+1)}$, and implies homogeneity (7) at the phase boundaries. A single phase implies Eq. (4a); an infinite number of continuous phases implies Eq. (4b) in the corresponding interval. The latter case implies a singularity at its boundary with the linear law interval (where $f'(v)$ is everywhere constant).

Thus, there exist two basic universal ways of "canonic" biological diversity and evolution: 1) Linear conservation law (which imposes no restrictions on the population heterogeneity in the values of canonic variables), and 2) Population homogeneity in the conserved values of canonic variables (which allows for any functional form of the universal law). The crossovers between different laws (linear–linear of linear–nonlinear) imply singularities in $v$.

Predicted linearity is experimentally explicit without any adjustable parameters, and allows for comprehensive verification. If in a certain interval a relation between two additive quantities is approximately linear for given populations, then it is also linear for any their mixtures (with the accuracy of the maximal deviation from linearity. One may significantly improve the accuracy of a linear approximation by discarding biologically or experimentally special cases for separate study). Empirical scaling[3, 5], which relates the basal oxygen consumption $v_e$ per maximal lifespan $e_m$ (for all animals) and $v_h$ per heartbeat $t_h$ (for animals with heart) to the animal mass $m$, is indeed approximately linear. This implies, in agreement with Eq. (4a), the predicted universal linear law for canonic fractions (here and on denoted by capital letters) $V_m$, $V_h$ vs $M$. The law is more explicit when it relates the numbers of basal oxygen atoms $N_m$ (consumed per maximal life span) and $N_h$ (consumed per heartbeat) to the number $N$ of body atoms:

$$N_m = AN, \quad N_h = BN. \tag{9}$$

This proves the existence of two universal biological constants (similar to fundamental physical constants, but known with ~30% accuracy): the numbers of consumed basal oxygen molecules per body atom per lifespan ($A \sim 12$) and per heartbeat ($B \sim 8 \times 10^{-9}$). In agreement with the first universality way, Eq. (9) allows for arbitrary heterogeneous populations (e.g., 2 elephants, 100 humans and 10 million hummingbirds).

Empirical scaling[2, 3, 5], which reduces basal oxygen consumption rate $\dot{v}_0$ per unit (i.e. biologically non-specific) time, heartbeat $t_h$ and respiration $t_r$ times, maximal lifespan $e_m$ to an animal mass $m$, is non-linear:

$$\dot{v}_0 = a_v m^\alpha; \quad t_h = a_h m^\beta; \quad t_r = a_r m^\gamma; \quad e_m = a_m m^\delta \tag{10}$$

where a's and "critical indexes" $\alpha$, $\beta$, $\gamma$, $\delta$ are constants[6]. Fluctuations of the variables in Eq. (10) (e.g., basal heartbeat rate in a given species, subspecies or breed) are relatively low, and any their heterogeneity in the population is inconsistent with Eq. (10). This agrees with the population homogeneity, predicted by the second way of universality in the case of a non-linear conservation law. Allometric relations (10) are closer for remote mammals and reptiles, or birds and amphibians, than for mammals and birds.[6] They modify the West et al laws[4] into more complicated relations (due to the impact of various biological and environmental factors). This is consistent with the unspecified functional form of a non-linear law.

Thus, conservation laws and their predictions are verified with metabolism, which is a must for entropy decrease, i.e. for survival. Now comprehensively verify them with mortality, which is a must for natural selection. Start with a test stone of human mortality, which is arguably the best quantified biological characteristic. Demographic "period" "life tables"[7] use accurately registered human birth and death records to calculate mortality rates $q_x$, i.e. the probabilities to die from age $x$ to $x+1$ in a given calendar year, for a given sex and country or its specific group (over 50 000 data items for Sweden alone). A "period" survivability $l_x$ is the probability to survive to a given age $x$ in a given calendar year. It equals $l_x = p_0 p_1 \ldots p_{x-1}$, where $p_y = 1 - q_y$ is the probability to survive from age $y$ to age $y+1$. "Cohort" life tables list $q_x$ and $l_x$ for a "cohort", born the same calendar year. Biodemographic life tables present mortality rates and survivabilities, usually for an animal cohort, at its characteristic ages (e.g., days for flies). Mortality is very sensitive to living conditions.[7-9] Yet, at any given age empirical relation between $q_x$ and $q_0 (= 1 - l_1)$ in protected populations is approximately universal and piecewise linear for species as remote as humans and flies.[10, 11] Survivability is additive, since the number of survivors in the population is the sum of their numbers in all population groups. This yields Eq. (2), where $u$, $v$ are replaced by canonic period survivabilities $L_x$ and $L_1$, but $f$ depends also on the "eigentime" $x$. According to demographic tables, mortality in the population is significantly heterogeneous in different groups with different living conditions. Then universality predicts linear dependence of $L_x$ on $L_1$. In fact, demographic data yield piecewise linear rather than linear dependence.[10, 11] This may be consistent with Eq. (6), which (in virtue of the dependence on $x$) changes to

$$L_x = a^{(j)}(x) L_1 + b^{(j)}(x) \text{ when } L_1^{(j)} \leq L_1 \leq L_1^{(j+1)}. \quad (11)$$

Verity predictions of Eq. (11). Linear conservation law was predicted to be universal for all animals. Indeed, when Eq. (11) is scaled according to its species specific crossovers, empirical $a^j(x)$ and $b^j(x)$ reduce to universal functions of age for species as remote as humans and flies.[10] Thus, the exact piecewise linear law of canonic survivability is biologically non-specific (i.e. independent of genotypes, phenotypes, life history, age specific diseases, and all other relevant factors). Some deceases, which significantly contribute to mortality (e.g., tuberculosis in pre-1949 Japan and in 1890-1940 Finland), do not violate the universal law.[11] So, a fraction of their mortality is also canonic.

At the crossovers and at the ultimate boundaries $L_1 = 0$ and $L_1 = 1$ of survival probabilities, Eq. (11) predicts homogeneous canonic survivability across the population. And indeed, heterogeneity of survivability decreases five fold to a minimum in the vicinity of the main crossover, then reaches a maximum, and finally decreases towards $l_1 = 1$[11], despite continuous monotonic improvement in living conditions (manifested by the fifty fold decrease in infant mortality). Thus, at the crossover susceptibility of survivability to different living conditions drastically decreases. Equation (11) also predicts that at any age crossovers occur at the same values of infant mortality (which correspond to different calendar years in different countries – e.g. 1949 in Sweden, 1963 in Japan, 1968 in France, 1970 in the USA), in agreement with ref. 10. Such crossovers are consistent with significant declines of old age mortality in the second half of the 20$^{th}$ century[12]. Demographers interpreted them as "epidemiological transitions", characterized primarily by the reduction of mortality from cardiovascular diseases. However, the predicted and verified rapid transitions occur simultaneously across generations (which have a different life history behind them and may even be genetically distinguishable). Piecewise linear dependence and crossovers are universal for species as remote as humans, med– and fruit flies.[10] This suggests that medical progress just shifts human survivability to a universal transition.

By Eq. (11), the "initial condition" (at $x=0$) $Q_0 = 1 - L_1$ accurately determines canonic mortality rate $Q_x = 1 - L_{x+1}/L_x$ at any age in the same calendar year. Mortality $Q_0$ strongly depends on living conditions, but from conception to $x=1$ only. So, at any age canonic mortality rate $Q_x$ rapidly adjusts to, and is determined by, current ($<2$ years for humans) living conditions only. It is independent of the previous life history. Therefore, together with $Q_0$, it may be rapidly reduced and reversed to its value at a much younger age. So, when mortality of a cohort is predominantly canonic, it may be reversed also. Reversible mortality implies its reversible adjustment to living conditions, and thus rapid accurate adaptability, with the relaxation time small compared to lifespan. Reduction of mortality $Q_x$ at any age to $Q_0$ is amazing (since living conditions, e.g., food and diseases, are intrinsically very different for elderly and newborns). Yet, it is consistent with clinical studies[13], as well as with demographic observation that infant mortality is a sensitive barometer of mortality at any age.[7] However, only exact universal law accurately predicts mortality reversibility, which is inconsistent with any evolutionary theory of aging.[9] Remarkably, this crucial prediction agrees with demographic data. For instance, mortality rate of Swedish females, born in 1916, at 48 years returned to its value at 20 years. Human survivability at any age extrapolates[11] to 1, suggesting that canonic mortality may be entirely eliminated. Thus, total mortality may be significantly decreases, and life expectancy significantly increased, in agreement with ref. 14 and other demographic data. In the last 30 years (1965-1995) Japanese females almost halved their mortality at 90 years, and increased their period probability to survive from 60 to 90 years 4.5-fold, to remarkable 33% of survivors. Neither this, nor vanishing susceptibility of survivability to living conditions, were anticipated in any of the theories[9], which relate mortality to mutation accumulation and cumulative damage; telomeres; oxygen consumption; free radicals; life-history trade-off; relation between reproductive rate and nutrient supply; and even lethal side-effect of a late-onset genetic disease.[15] In contrast to these theories, universal conservation law suggests the existence of a new unusual mechanism of mortality, which allows for rapid reversible adjustment to changing living conditions (in particular via singularities and population homogenization), and dominates in evolutionary unprecedented protected populations. The only known reversible processes in a macroscopic system are adiabatic changes in its equilibrium state. Exact universal (i.e., biologically non-specific, independent of genotypes, phenotypes, life history, age specific diseases, and all other relevant factors) law is characteristic for physics rather then biology. Thus, its accurate mapping (8) onto the equilibrium of universal phases, its singularities, and its number of variables (unlike multi-variable potential in the conservation law of mechanical energy) may not be just a coincidence. Validity of unanticipated predictions of the conservation laws calls for their comprehensive study in quantitatively controllable conditions (such as temperature, humidity, pressure and oxygen concentration in the air, etc). One may, e.g., study metabolism and mortality in genetically homogeneous cohorts of different ages as a function of sufficiently slowly and non-monotonically changing temperature. When the temperature in one of the cohorts does not change, and in another slowly changes and then returns to its initial value, one may verity mortality reversibility. Among different parameters one may chose those whose change allows survivability to reach the crossover. There the mortality change is predicted to be the slowest. The scaling for bacteria (with fission time replacing life span) is close to the universal scaling for the life span and oxygen consumption. This suggests a study of bacteria fission, as well as their metabolic and dynamic characteristics, which may grossly simplify the search for molecular nature of universality and for a biologically non-specific "pill" to regulate it.

To summarize. Conservation laws, i.e. exact relations between certain (dominant) fractions of biological quantities, are derived. They reduce to universal biological and evolutionary constants, similar to fundamental constants in physics, and predict "quantized" species specific constants. Singularities in the laws yield vanishing susceptibility to different conditions The laws specify perspectives of and impose limitations on biological diversity and its evolutionary changes. Perspectives include a possibility to rapidly and accurately direct adaptability and decrease mortality, perhaps even with a biologically non-specific pill. The laws, their unanticipated implications and predictions are verified with metabolism, which is a must for entropy decrease, and thus for survival, and with mortality, which is a must for natural selection.

Acknowledgement. Financial support from A. von Humboldt award and R. & J. Meyerhoff chair is appreciated. I am very grateful to I. Kolodnaya for technical assistance.

[+] Permanent address

## References

1. Consider an arithmetic cartoon of irreversibility. Multiply $1/7$ by 8; take the fractional part of the product; repeat the procedure – it always ends up with the "reversible" $1/7$. However, approximate $1/7$ with 0.14, 0.143, 0.14286, etc. The results are the further off from $1/7$, the better is the approximation. This game is directly related to chaos, entropy increase, and the very foundations of statistical mechanics.

    Note that biological systems are an evolutionary miracle. Reproduction allowed them to meet uniquely low chances to survive in natural selection. After only few million generations these chances are 1 out of $10^{-1,000,000}$.

2. Schmidt-Nelsen, K. Scaling: Why is Animal Size so Important? (Cambridge Univ. Press, Cambridge, 1984); Calder III, W.A.. Size, Function and Life History (Harvard Univ. Press, Cambridge, MA, 1984); Clark, A.J. Comparative Physiology of the Heart (Macmillan, N.Y., 1987).

3. Azbel', M.Ya. *PNAS USA* **91**, 12453 (1994).

4. West, G.B, Woodruff W.H., Brown J.H. *PNAS USA* **99**, 2473 (2002), Gillooly, J. F., Charnov, E. L. West, G. B. Savage, V. M. Brown, J. H., Nature, **417**, 70 (2002), and refs. therein.

5. Oxygen consumption of cold-blooded animals exponentially depends on temperature.[2-4] (Their activation energy is consistent with the hydrogen binding energy[3]). When it is renormalized to $311°K$, then it yields the same linear universal law as for warm-blooded animals.

6. The estimates for "critical indexes" $\alpha$, $\beta$, $\gamma$, $\delta$ vary[1-3] from 2/3 to 3/4 for $\alpha$, from 1/4 to 1/3 for $\beta$, $\gamma$, $\delta$. It was suggested[3] that critical indexes in Eq. (10) are related to fractality of, e.g., blood vessels in a body. In a model of fractal – like distribution networks, West et al.[4] derived $\alpha = 3/4$. To compare different animal classes, it is convenient to present them in the dimensionless form.[3] For instance, relate the number of consumed oxygen molecules $\dot{n}_0$ per second to the number of body atoms $n$: $\dot{n}_0 = an$, For mammals $\alpha = 0.76 \pm 0.01$; $a = 0.1 \pm 0.02$; for birds $\alpha = 0.7 \pm 0.01$; $a = 4.5 \pm 1$; for reptiles $\alpha = 0.73 \pm 0.02$; $a = 0.12 \pm 0.03$; for amphibians $\alpha = 0.66 \pm 0.04$; $a = 5.5 \pm 2$; for fishes $\alpha = 0.81 \pm 0.02$; $a = 0.002 \pm 0.0008$; for invertebrates $\alpha = 0.79 \pm 0.03$; $a = 0.002 \pm 0.0008$.

7. Lee, R.D. & Carter, L.R. *J. Am. Stat. Assoc.* **87**, 659 (1993); Coale, A.J.,Demeny, P., Vaughan, B., Regional Model Life Tables and Stable Populations. 2nd ed., Acad. Press,N.Y.,1993, and refs. therein.

8. For instance, human mortality changes by two orders of magnitude with country and time[7], genotypes and life history[9], population heterogeneity- Carnes, B.A. & Olshansky, S.J. *Exp. Gerontol.* **36**, 447 (2001), epidemics, age specific factors and deceases, acquired components (also for genetically identical populations in the same environment[9]), and even the month of birth- Doblhammer, G. & Vaupel, J.W. *PNAS USA* **98**, 2934 (2001).

9. Finch, C.E. Longevity, Senescence and Genome. Chicago Univ. Press, Chicago (1990); Finch, C.E. & Kirkwood, T.B. Chance, Development and Aging. Oxford Univ. Press, Oxford (2000); Charlesworth, B. Evolution in Age-Structured

Populations. Cambridge Univ. Press, Cambridge (1994).; Partridge, L. & Gems, D. *Nature Rev. Genet.* **3**, 165 (2002).

10. Azbel' M.Ya. *PNAS USA* **96**, 3303 (1999).

11. Azbel', M.Ya. *Exp. Geront.* **37**, 859 (2002) and refs. therein.

12. Wilmoth, J.R., Deagan, L.J., Lundstron, H. & Horiuchi, S. *Science* **289**, 2366 (2000); Tuljapurkar, S., Li, N. & Boe, C. *Nature* **405**, 729 (2000), and refs. therein.

13. Osmond, C. & Barker, D.J.P. *Envir. Health Persp.* **108**, 543 (2001).

14. Deppen, T. & Vaupel, J.W. *Science* **296**, 1029 (2002).

15. Partridge, L. & Gems, D. *Nature* **418**, 921 (2002).